\documentclass[11pt]{article}
\usepackage{times}
\usepackage{geometry}
\usepackage[utf8]{inputenc}
\usepackage{enumitem,amssymb}
\usepackage{graphicx}
\usepackage{fancyhdr}
\usepackage{aas_macros}
\usepackage{mdframed} 
\usepackage{xcolor}
\usepackage{eurosym}

\usepackage{hyperref}
\hypersetup{
    colorlinks=true,
    linkcolor=blue,
    filecolor=blue,      
    urlcolor=blue,
}
\mdfdefinestyle{theoremstyle}{
innertopmargin=\topskip,}
\mdtheorem[style=theoremstyle]{lrptextbox}{}

\mdfdefinestyle{recstyle}{leftmargin=0.5cm,rightmargin=0.5cm,%
innerleftmargin=0.2cm,innerrightmargin=0.2cm,roundcorner=10pt}

\mdfdefinestyle{execstyle}{leftmargin=0.5cm,rightmargin=0.5cm,%
innerleftmargin=0.4cm,innerrightmargin=0.4cm,roundcorner=10pt}

\newcommand{\vs}{\vspace{5pt}}
\newcommand{\vsneg}{\vspace{-5pt}}
\newcommand{\no}{\noindent}

\begin{document}

\vs\vs\vs\vs

\begin{center}
   {\bf \sc \LARGE Canada and the SKA from 2020~--~2030}
\end{center}

\vs\vs\vs

\parbox{0.92\textwidth}{%
Kristine Spekkens (RMC; Canadian SKA Science Director), Cynthia Chiang (McGill) , Roland Kothes (NRC), Erik Rosolowsky (Alberta), Michael Rupen (NRC),
Samar Safi-Harb (Manitoba), Jonathan Sievers (McGill), Greg Sivakoff (Alberta), Ingrid Stairs (UBC), Nienke van der Marel (NRC), Bob Abraham (Toronto), Rachel Alexandroff (Toronto), Norbert Bartel (York), Stefi Baum (Manitoba), Michael Bietenholz (York), Aaron Boley (UBC), Dick Bond (Toronto), Joanne Brown (Calgary), Toby Brown (McMaster), Gary Davis (SKAO), Jayanne English (Manitoba), Greg Fahlman (NRC), Laura Ferrarese (NRC), James Di Francesco (NRC), Bryan Gaensler (Toronto), Severin Gaudet (NRC), Vanessa Graber (McGill), Mark Halpern (UBC), Alex Hill (UBC), Julie Hlavacek-Larrondo (Montreal), Judith Irwin (Queen's), Doug Johnstone (NRC), Gilles Joncas (Laval), Vicky Kaspi (McGill), JJ Kavelaars (NRC), Adrian Liu (McGill), Brenda Matthews (NRC), Jordan Mirocha (McGill), Raul Monsalve (McGill), Cherry Ng (Toronto), Chris O'Dea (Manitoba), Ue-Li Pen (CITA), Rene Plume (Calgary), Tim Robishaw (NRC), Sarah Sadavoy (Queen's), Viraj Sanghai (Dalhousie), Paul Scholz (NRC), Luc Simard (NRC), Richard Shaw (UBC), Saurabh Singh (McGill), Kris Sigurdson (UBC), Kendrick Smith (Perimeter), David Stevens (MDA), Jeroen Stil (Calgary), Sean Tulin (York), Cameron van Eck (Toronto), Jasper Wall (UBC), Jennifer West (Dunlap), Tyrone Woods (NRC), Dallas Wulf (McGill)
    }

\vs\vs\vs\vs\vs 
{\Large \bf Executive Summary}
   
\vs\vs  
   
   The Square Kilometre Array (SKA), an exciting new world observatory that will enable transformational science at metre and centimetre wavelengths for years to come, is rapidly becoming reality. The SKA will be built in two phases, with the first phase (SKA1) representing ~10\% of the full facility (SKA2). The SKA1 Design Baseline development is almost complete, and construction is set to begin early in the next decade.  When constructed, it will be the largest and most powerful general-purpose radio telescope operating from 50 MHz -- 15 GHz for years to come. Scientific and technological participation in the SKA has been identified as a top priority for the Canadian astronomical community for almost twenty years. This 2020 Canadian Long-Range Planning process (LRP2020) white paper advocates for Canada’s continued scientific and technological participation in the SKA project, focussing on Canadian prospects for SKA1 from 2020~--~2030. 

\vs

SKA1 is poised to make fundamental advances across a broad range of fields by virtue of its combination of sensitivity, angular resolution, imaging quality and frequency coverage. SKA1 scientific goals align well with the strengths of Canadian researchers.  Canada is a world leader in studies of pulsars, cosmic magnetism and transients, as well as in low-frequency cosmology.  Our multi-wavelength expertise in galaxy evolution, multi-messenger astronomy and planetary system formation –- in which radio observations play a critical role –- is also a key strength. The Canadian community therefore has the potential to carry out important PI science with SKA1, as well as to play world-leading roles in a number of the transformational Key Science Projects (KSPs) that are anticipated to take up the majority of available telescope time. An examination of the KSP leadership opportunities afforded by a decade of full operations implies that a 6\% participation in the SKA1 Design Baseline is well-matched to Canadian scientific capacity and ambitions.

\vs

Canada is a leader in technological development for the SKA through effective partnerships between universities, the National Research Council (NRC) and industry.  Our key SKA1 technological capabilities include the design and fabrication of correlators and beamformers, digitisers, low-noise amplifiers, signal processing, and monitor \& control.  These technologies provide a suite of possible in-kind contributions to offset construction costs for good return on the capital investment required to participate in SKA1 at a level commensurate with our scientific ambitions. Canada also has the computing platform and archive development expertise to make important contributions to the SKA Regional Centre (SRC) network that will deliver global SKA1 scientific computing capability. A Canadian SRC would leverage our national compute strength to provide processing, storage, and user support tailored to Canadian SKA1 needs while also fulfilling our SKA1 participation requirements.

\newpage

Canadian contributions to the SKA now span two decades, marked by scientific and technological leadership that persists today within a vibrant metre and centimetre-wave radio community. SKA1 is happening now. Canada at last has the opportunity to reap the scientific benefits of our contributions, while an early commitment to construction would maximize our impact on this phase and our technological benefits as well.

\vs

LRP2020 will determine the future of the SKA in Canada for the next decade and beyond. We make the following recommendations:

\vs\vs

\begin{mdframed}[style=execstyle]

\vs

\no {\bf 1. Canada should participate in the construction and operations phases of SKA1.} SKA1 Design Baseline construction, operations and a staged technology development program should be funded at a 6\% level, commensurate with Canadian scientific ambitions. This commitment is estimated to cost \$160M CAD (2017 CAD) over the period 2021~--~2030.

\vs

\no { \bf 2. Canada should participate in the SKA Regional Centre (SRC) network to ensure community access to the processing, storage and user support required to scientifically exploit SKA1.} The cost of this participation at a level commensurate with Canadian scientific ambitions, and in accordance with SRC network guidelines, is estimated to be \$45M CAD (2017 CAD) over the period 2021~--~2030 in addition to construction and operations funding. To meet its SKA1 compute needs, Canada should leverage its established strength in scientific computing platforms and archive development by hosting a Canadian SRC.

\vs

\no {\bf 3. The membership model through which Canada participates in the intergovernmental organisation (IGO) that will build and operate SKA1 should provide full scientific and technological access as well as leadership rights for Canadian researchers and industry.}  An agreement for Canadian participation in the IGO should be finalized as early as possible in the next decade in order to maximize our impact on the construction phase as well as to maximize opportunities for technological tender and procurement.

\vs

\end{mdframed}

\newpage

\vs
\begin{center}
   
   \vs\vs\vs
   {\bf \sc \LARGE Canada and the SKA from 2020~--~2030}
   \vs\vs\vs

\parbox{0.92\textwidth}{%
\bf \hspace{10pt} This white paper submitted for the 2020 Canadian Long-Range Planning process (LRP2020) presents the prospects for Canada and the Square Kilometre Array (SKA) from 2020~--~2030, focussing on the first phase of the project (SKA1) scheduled to begin construction early in the next decade. SKA1 will make transformational advances in our understanding of the Universe across a wide range of fields, and Canadians 
are poised to play leadership roles in several. Canadian key SKA technologies will ensure a good return on capital investment in addition to strong scientific returns, positioning Canadian astronomy for future opportunities well beyond 2030. We therefore advocate for Canada's continued scientific and technological engagement in the SKA from 2020~--~2030 through participation in the construction and operations phases of SKA1.
    }
\end{center}

\section{SKA1 Overview \label{sec:project}}

The Square Kilometre Array (SKA) is one of the most ambitious astronomy projects on the horizon today, with broad scientific goals and challenging technical requirements. The SKA will be built in two phases, with the first phase (SKA1) representing $\sim$10\% of the full facility (SKA2).  When constructed, SKA1 will be the largest and most powerful general-purpose radio telescope operating from 50$\,$MHz -- 15$\,$GHz for years to come.  

\vs

\begin{figure*}[t]
\begin{center}
\includegraphics[width=0.9\textwidth]{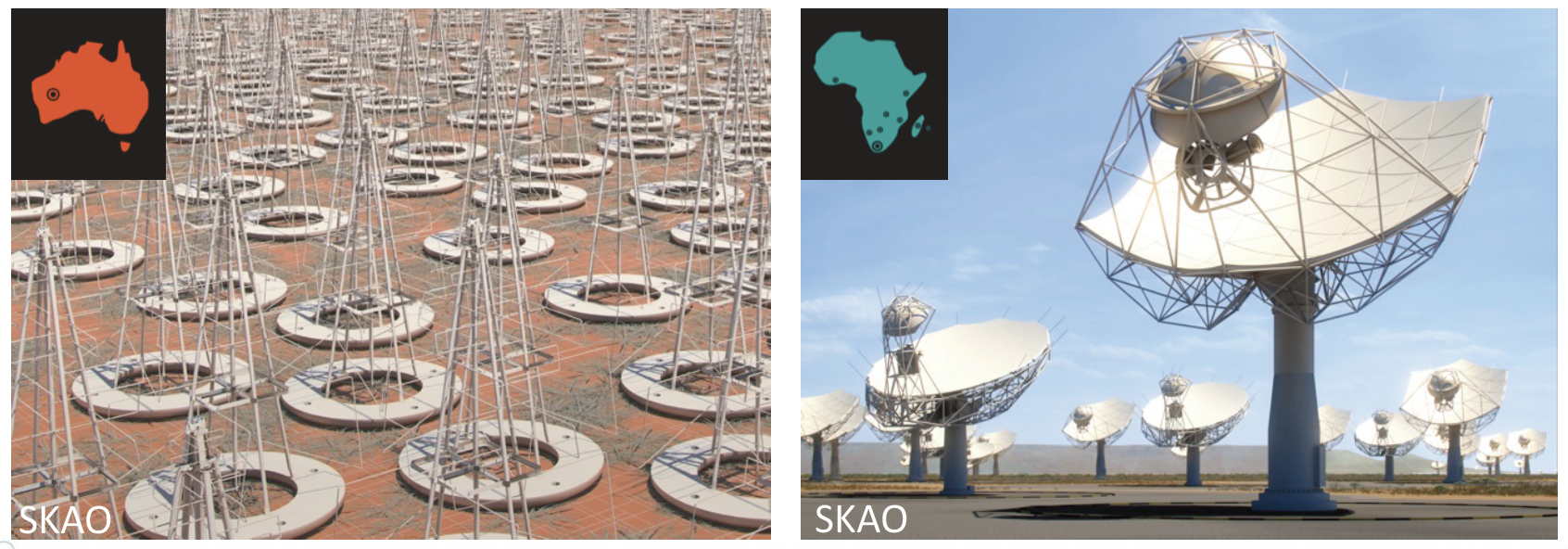}
\caption{Artist's conception of SKA1-Low antennas (left panel) and SKA1-Mid dishes (right panel) with approximate locations in Australia and South Africa (also showing African VLBI stations) inset. Image credit: SKAO.  \label{fig:SKAartist}}
\end{center}
\end{figure*}
\begin{figure*}[t]
\begin{center}
\includegraphics[width=0.77\textwidth]{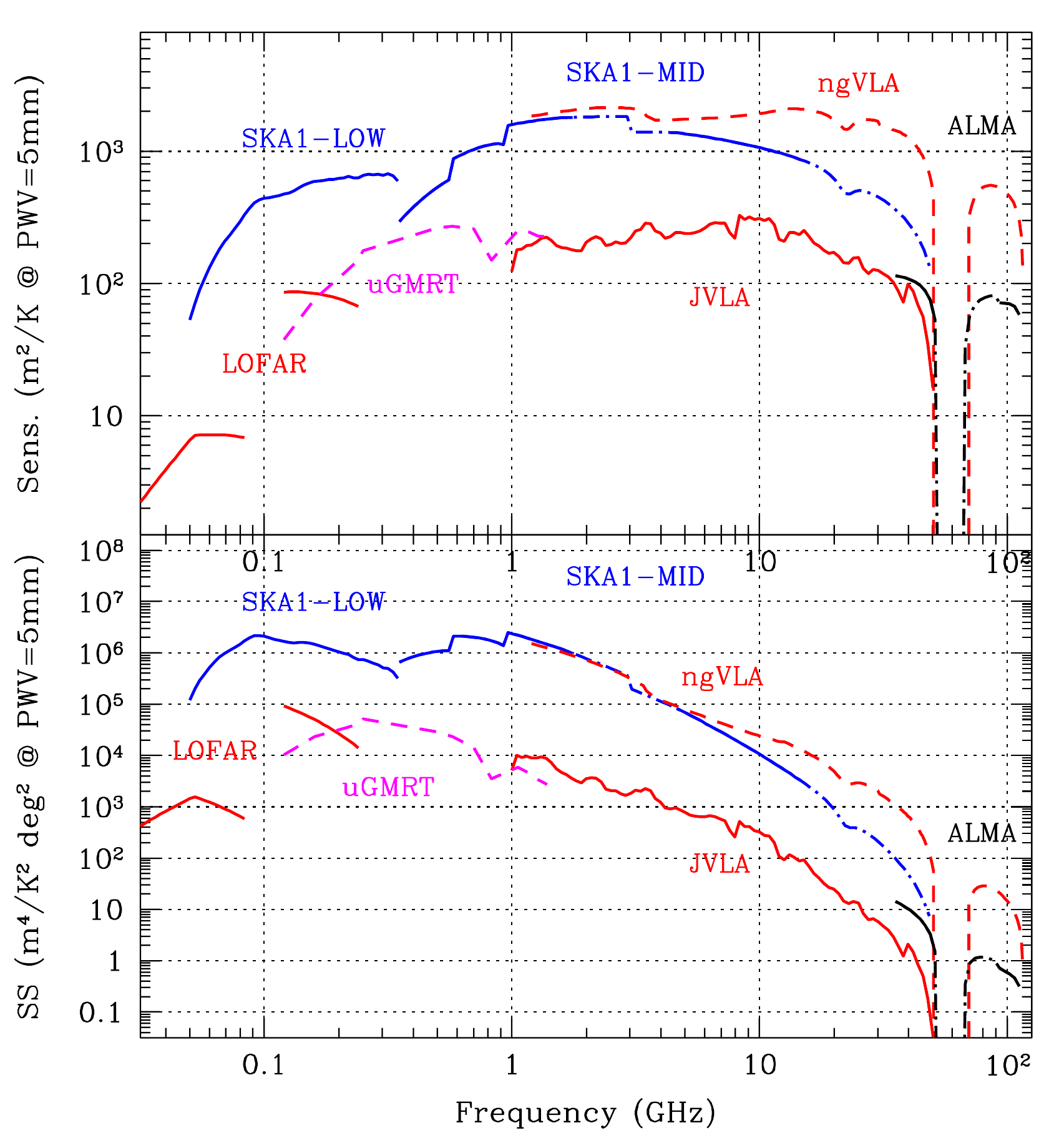}
\caption{Anticipated SKA1 sensitivity (top) and survey speed (bottom) as a function of frequency in comparison to existing (LOFAR, JVLA, ALMA) and future (uGMRT, ngVLA) imaging facilities. Image credit: SKAO. \label{fig:sens}}
\end{center}
\end{figure*}

The SKA1 Design Baseline is now mature, and represents the construction and operations benchmark for the next decade. SKA1 will consist of two facilities: SKA1-Low (50$\,$MHz -- 350$\,$MHz) located in Australia, and SKA1-Mid (350$\,$MHz -- $15+$\,GHz) located in South Africa (Fig.~\ref{fig:SKAartist}). The SKA1-Low Design Baseline comprises 512 stations each with 256 antennas, providing continuous frequency coverage from 50$\,$MHz -- 350$\,$MHz and a maximum interferometric baseline of 65$\,$km (Dewdney+ 16). Averaged over the LOFAR band, SKA1-Low will have 1.25 times the LOFAR resolution, 8 times its sensitivity and 135 times its survey speed (see Fig.~\ref{fig:sens}).  The SKA1-Mid Design Baseline comprises $133 \times 15\,$m dishes as well as the $64 \times 13.5\,$ dishes from the MeerKAT SKA Pathfinder working in concert from 350$\,$MHz -- $15+$\,GHz, with a maximum interferometric baseline of $150\,$km (Dewdney+ 16). The frequency range spanned by SKA1-Mid is divided into several bands, of which 3 are part of the Design Baseline: Band~1 will contiguously span $350\,$MHz -- $1.05\,$GHz, Band~2 will contiguously span $950\,$MHz -- $1.76\,$GHz, and Band 5 will span $4.6\,$GHz -- $15.3\,$GHz in $2 \times 2.5\,$GHz sub-bands.  When averaged over the overlapping (J)VLA bands,  the SKA1-Mid Design Baseline will have 4 times the resolution of the (J)VLA in its most extended (``A") configuration, 4 times its point-source sensitivity and 60 times its survey speed (see Fig.~\ref{fig:sens}). 

\vs

SKA1 is poised to make fundamental advances across a broad range of fields by virtue of its combination of sensitivity, angular resolution, imaging quality and frequency coverage. In particular, SKA1 will be a superb survey instrument (see Fig.~\ref{fig:sens}). While the operations model for SKA1 will be finalized during the construction phase, the majority of the available observing time, $50\% - 75\%$, is expected to be earmarked for the execution of large programs called Key Science Projects (KSPs) by participating countries, compared to $25\% - 45\%$ for principal-investigator (PI) programs by participating countries and $\lesssim 5\%$ open time. The scientific legacy of SKA1 is therefore likely to lie in the suite of KSPs ultimately carried out.

\vs 

\begin{figure*}[t]
\begin{center}
\includegraphics[width=0.95\textwidth]{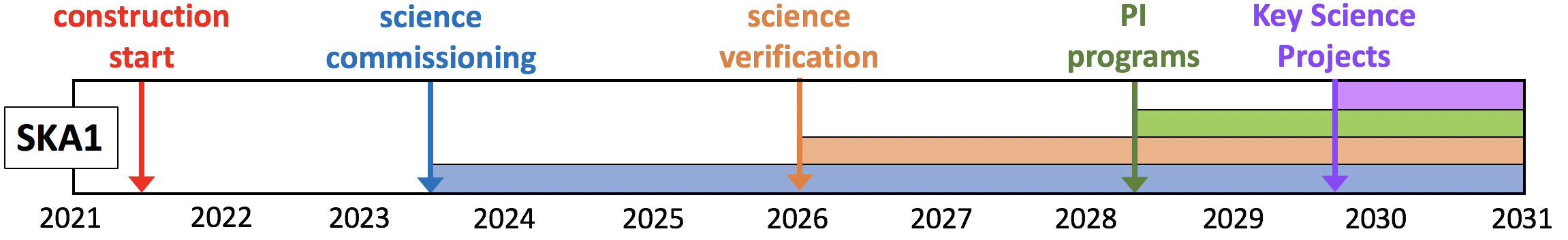}
\caption{Projected SKA1 scientific milestones during construction, assuming a construction start in Q2~2021.  \label{fig:timeline}}
\end{center}
\end{figure*}

With a mature Design Baseline and the critical design review (CDR) process for the individual design elements and the integrated system nearly complete (SKAO+ 19), the construction timeline for SKA1 is becoming more concrete. A representation of that timeline which focusses on scientific milestones is shown in Fig.~\ref{fig:timeline}. The timeline is anchored by an anticipated construction start in Q2~2021 and a nominal construction duration of 7 years. Science commissioning will start as soon as the first dish or station is on site in 2023, while science verification with antenna assemblies that are scientifically competitive with current facilities (64 SKA1-Low stations, 64 SKA1-Mid dishes) is anticipated in 2026. Shared-risk observations with the full complement SKA1-Mid and SKA1-Low arrays are expected to begin in late 2027. Full SKA1 operations are expected in 2028, with PI projects scheduled towards the end of that year and KSPs on the telescope in late 2029. The SKA1 construction and ramp-up to full operations will therefore be complete in the next decade.

There are 3 basic elements to the cost of SKA1 (excluding science compute; see below) in the next decade: 1) the cost of procuring and constructing the facilities in Australia and South Africa, 2) the cost of construction activities at SKA Headquarters in the UK, and 3) the cost of early operations and business-enabling functions ramping up to full operations. This white paper will follow the historical project convention of defining ``construction" as the first element, and ``operations" as the second two. The most recent estimated construction cost of the Design Baseline is \euro$\,$914M (2017 Euros; C\'esarsky 19b), while operations costs through the end of 2030 are 
comparable to the construction cost (C\'esarsky 19a). It is important to note that the Design Baseline and operations costing will soon evolve, with a bottom-up Cost Book using the same methodology as other large projects (e.g.\ the TMT) anticipated in Q2~2020. This costing will also define the final ``Deployment Baseline" of SKA1, which corresponds to as much of the Design Baseline as can be afforded for a construction cap of \euro$\,$691M (2017 Euros; Womersley 13, C\'esarsky 19b). Since the well-defined Design Baseline is the construction benchmark, and since 
the scalable nature of interferometers makes it relatively straightforward to recover from any cost-saving cuts,
the ambitions for the Canadian scientific and technological community for SKA1 in this white paper reference the Design Baseline. 

\vs

The raw data rates, processing speeds and calibrated data volumes implied by the SKA1 Design Baseline are enormous, leading to the adoption of a tiered model for data and science support. Similar to the model adopted by CERN, a network of SKA Regional Centres (SRCs) will handle the global science processing, archive and user support needs for SKA1. In this white paper, the anticipated cost for Canadian SRC contributions is considered separately from construction and operations costs (\S\ref{sec:rec}).

\vs

SKA Design and pre-construction activities in the last decade have been overseen by the SKA Organisation (SKAO), a UK-based company which currently comprises 13 participating countries: Australia, Canada, China, France, Germany, India, Italy, New Zealand, South Africa, Spain, Sweden, The Netherlands, and the UK.  In the SKA1 construction and operations phases, stewardship of the project will transition to an intergovernmental organisation (IGO) called the SKA~Observatory. An IGO, which is the governance structure that underlies CERN and ESO, differs fundamentally from a company in that it provides sovereign protection to the organization and its employees. The 7 countries that signed the IGO Treaty Convention and Final Record in March 2019 (Australia, China, Italy, South Africa, The Netherlands, and the UK) are the Founding Members of the IGO, which becomes active once the Convention is ratified by 5 of them. The Netherlands became the first country to ratify the Convention in August 2019, and the IGO is on pace to come into force by Q3~2020. 

\vs 

The most significant governance milestone in the history of the SKA project is the upcoming transition from the SKAO and its Board of Directors to the IGO and its Council, which is currently anticipated for Q4~2020 (C\'esarsky~19b). The last major goals of the SKAO are to deliver detailed costing, construction and operations proposals to the IGO Council for their approval in Q3~2020. The Council Preparatory Task Force (CPTF) represents the interests of the IGO until the Convention is ratified, and the focus of CPTF activities is the development of a funding schedule, the establishment of principles governing Associate IGO Membership, and the finalisation of key documentation such as procurement strategy and IP policy (C\'esarsky~19a).  Since funding commitments towards SKA1 construction and operations were decoupled from treaty negotiations early in the process, the IGO will exist before financial commitments are in place. The first major task of the IGO Council at its inaugural meeting will be to approve a budget for the IGO's first year of operations (anticipated to be 2021; C\'esarsky~19b).

\section{Canada and the SKA \label{sec:Canada}}

Canada has a 20+ year history of significant scientific and technological contributions to the SKA initiative, and Canadian governance leadership was instrumental in its early days (Ekers 12). A detailed account of the history and current status of Canada and the SKA is given in the final report submitted to the LRP~2020 panel (Spekkens+ 19); this section focusses on science, technology and governance in Canada pertaining to its future participation in SKA1 construction and operations.

\begin{figure*}[t]
\begin{center}
\includegraphics[width=0.9\textwidth]{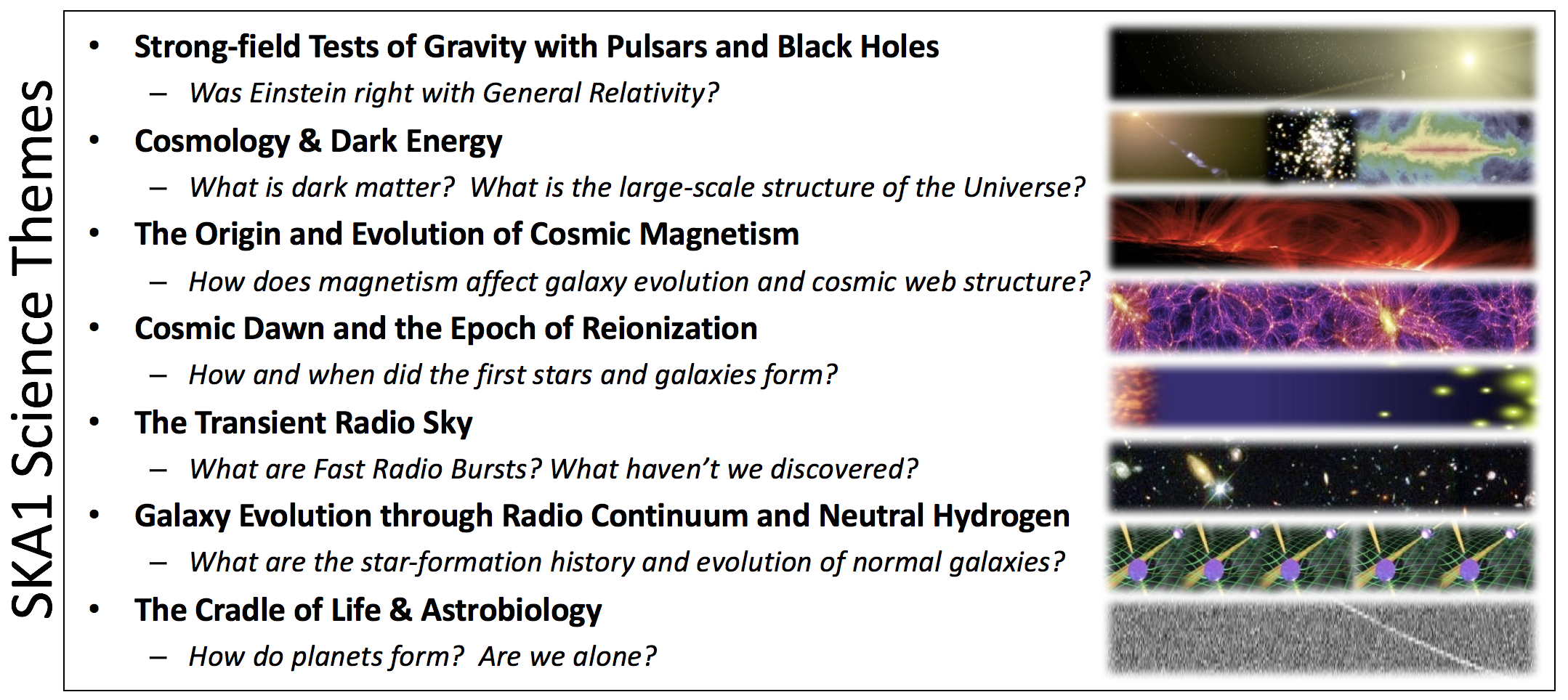}
\caption{SKA1 science themes (Braun+ 15). 
Image credits (from top): NASA; NASA/StSci/AURA/MPIFR/G. Haslam; LMSAL/NASA; Springel+ 05; SKAO; NASA/ESA/StSci; MPIFR/M. Kramer.   \label{fig:science}}
\end{center}
\end{figure*}

The science goals of SKA1 are among the broadest of any current or planned facility worldwide (SKAO+ 15), and are represented by the themes and questions given in Fig.~\ref{fig:science}. They result from
a community-based exercise to produce Highest-Priority Science Objectives (HPSOs; Braun+ 14) that informed the SKA1 Level~0~Science Requirements (Braun+ 15), as well as emerging fields such as fast radio bursts and gravitational waves. Fundamentally answering many of these questions is expected to require large allocations of observing time -- $5000$ to $10,000\,$hours -- in order to exploit SKA1's survey capabilities (Braun+ 15; see also Fig.~\ref{fig:sens} and \S\ref{sec:project}). It is therefore reasonable to expect that the KSPs that will define the scientific legacy of SKA1 (\S\ref{sec:project}) will focus on these themes.

\vs

The SKA1 scientific goals embodied by the themes and questions in Fig.~\ref{fig:science} align well with the strengths of Canadian researchers. Canadians are members of all 11 SKA Science Working Groups (SWGs), 
and have co-chaired 4 of them in the past 5 years.  In particular, Canada is a world leader in studies of pulsars, cosmic magnetism and transients, as well as in low-frequency cosmology.  Our multi-wavelength expertise in galaxy evolution, multi-messenger astronomy and planetary system formation – in which radio observations play a critical role – is also a key strength. Details regarding the particular advances that SKA1 is expected to make in these fields as well as the capacity for Canadian leadership in each one are summarized in Boxes~1~and~3 at the end of this white paper, as well as in several others submitted for LRP~2020 (e.g.\ Kaspi+ 19; Fonseca+ 19; Liu+ 19a,b; Ruan+ 19; Sadavoy+ 19; Stairs+ 19; West+ 19). The Canadian community therefore has the potential to carry out important PI science with SKA1, as well as to play world-leading roles in a number of transformational KSPs.

\vs

The high prioritization of SKA R\&D in previous Long-Range Plans and reviews 
 has helped forge and maintain effective partnerships between government, academia and industry for Canadian SKA1 technology development (Rupen+ 19; Spekkens+ 19). Canadian leadership in SKA1-related technologies includes the design and fabrication of correlators and beamformers for SKA1-Mid, low-noise amplifiers and digitisers for SKA1-Mid Bands~1~and~2, and signal processing and monitor~\&~control for SKA1-Mid. Notably, Canada led the Central Signal Processor element consortium that passed CDR with no action - the only consortium to have received this high rating. The technologies above provide a suite of possible in-kind contributions to offset SKA1 construction costs for good return on capital investment. Canada also has the computing platform and archive development expertise to make important contributions to the SKA Regional Centre (SRC) network that will deliver global SKA1 scientific computing capability. Further details regarding Canadian technological leadership and return on investment for SKA1 are given in Boxes~3,~5~and~6 of this white paper.

\vs 

Canada currently contributes to the governance of the SKA project through its membership in the SKA Organisation (SKAO), with the National Research Council (NRC) as the adhering organisation (see Box~4). However, the anticipated transition in global project leadership from the SKAO to the intergovernmental organisation (IGO) for construction and operations (C\'esarsky+19b; \S\ref{sec:project}) implies that the SKAO's governance role in SKA1 is becoming increasingly limited. Canadian participation in the IGO is complicated by laws stipulating that governmental approval must be obtained before treaty negotiations can be entered. NRC has been directed by the Canadian government to explore IGO membership options, and Canada is now an Observer country to the Council Preparatory Task Force (CPTF). Bilateral discussions with the CPTF regarding possible SKA1 participation mechanisms for Canada are ongoing.




\section{Canadian Participation in SKA1: Recommendations \label{sec:rec}}

SKA1 is on the verge of construction (\S\ref{sec:project}), and its scientific goals and technical challenges align well with Canadian strengths (\S\ref{sec:Canada}).  Considerations for Canada's future participation in SKA1 and recommendations for 2020~--~2030 are below. 

\vs

Several lines of reasoning suggest that a participation level in the SKA1 Design Baseline of $6\% \pm 2\%$ is appropriate for the Canadian community.  Focussing on science, the total number of KSP leadership positions over a decade has been calculated given current estimates of the SKA1 operations model, a distribution of KSP durations commensurate with SKA1 scientific goals and a KSP management structure similar to that of current SKA Pathfinder surveys (c.f.\ Appendix A of Spekkens+ 19). These numbers are preliminary since telescope operations and KSP structure have yet to be finalized, but they are the most sophisticated ones currently available across the project and therefore a reasonable benchmark. This model suggests that a 6\% participation level affords leadership of $\sim$ 4 medium/large (ie. 5000--10000 hours) KSPs, chairship of $\sim$40 KSP working groups (WGs), and $\sim300$ KSP team memberships over the course of a decade. These numbers are broadly consistent with the expertise and ambitions from \S\ref{sec:Canada}: Canadians could lead a KSP in each of pulsars, magnetism, transients and cosmology, lead $\sim$5 KSP WGs in each of pulsars, magnetism, transients, cosmology, galaxies, GWs and planets, and have a few team memberships in all KSPs project-wide.

\vs

If Canada does participate in the SKA1 Design Baseline at the 6\% level, then the latest available costing and schedule imply that construction funding of order (\euro$\,918\mathrm{M} \times 0.06 \times 1.46\,\mathrm{CAD/}$\euro$\,\sim$) \$80M$\,$CAD (2017~CAD) will be required in the next decade, as well as commensurate operations funding for a total estimated construction and operations cost of \$160M~CAD (2017 CAD) from 2021~--~2030. This funding includes contributions to a staged observatory development fund that grows to a project-wide \euro$\,20\mathrm{M}$/year (2017 \euro) by the onset of full operations.   

\vs

\begin{mdframed}[style=recstyle]
        {\bf Recommendation 1. Canada should participate in the construction and operations phases of SKA1.} SKA1 Design Baseline construction, operations and a staged technology development program should be funded at a 6\% level, commensurate with Canadian scientific ambitions. This commitment is estimated to cost \$160M~CAD over the period 2021~--~2030.
\end{mdframed}

\newpage

In addition to construction and operations funding, Canada will need to contribute to the global SKA Regional Centre (SRC) network that will supply the requisite processing, storage and user support to scientifically exploit SKA1 (see \S\ref{sec:project}). The estimated Canadian SRC cost in the next decade during construction and ramp up to full operations has been calculated from the bottom up (c.f.\ Appendix~B of Spekkens+ 19), and is the most sophisticated currently available across the project: it includes PI and KSP networking, processing and online storage, near-line storage for the SKA1 archive in compliance with global SRC guidelines (Bolton+ 18), 5-year processing refresh  and continuous storage refresh, and staffing to provide user support for both observers and archive users. 

\vs

For a 6\% Canadian participation level in the SKA1 Design Baseline, the estimates described above predict that the SRC cost will be \$45M~CAD (2017~CAD) from 2021~--~2030, where the bulk of the cost stems from on-line storage, followed by support staff, processing and networking. The first hardware would be purchased and minimal processing/storage carried out in 2022, with processing/storage/support capacity ramping up to handle full SKA1 Design Baseline operations by 2028.  A Canadian SRC (as opposed to a model in which science compute is purchased elsewhere) would leverage Canadian leadership in scientific computing platform and archive development to provide processing, storage, and user support tailored to Canadian SKA1 needs. SRC capabilities are part of the broader digital research infrastructure in astronomy that will be required in the next decade (Kavelaars+ 19).

\vs

\begin{mdframed}[style=recstyle]
{\bf Recommendation 2. Canada should participate in the SKA regional centre (SRC) network to ensure community access to the processing, storage and user support required to scientifically exploit SKA1.} The cost of this participation at a level commensurate with Canadian scientific ambitions, and in accordance with SRC network guidelines, is estimated to be \$45M~CAD over the period 2021~--~2030 in addition to construction and operations funding. To meet its SKA1 compute needs, Canada should leverage its established strength in scientific computing platforms and archive development by hosting a Canadian SRC.
\end{mdframed}

\vs



SKA1 is happening, and Canada at last has the opportunity to reap the scientific benefits of  decades-long technical and governance contributions by participating in the construction and operations phases. In order to do so, a mechanism for participating in the IGO that will soon take over the project is needed. SKA1 construction tender and procurement will take place soon after the construction phase begins in 2021 (\S\ref{sec:project}); a participation agreement finalized early in the next decade would maximize Canadian opportunities to secure construction contracts as well as to inform policy and process during the construction phase.   

\vs

It is possible for Canada to accede to the IGO if governmental approval is obtained; alternatively, Canada could participate in SKA1 construction and operations via some form of Associate Membership or other agreement to be negotiated with the IGO. The IGO Treaty Convention and Final Record do not distinguish between the scientific or technological access and leadership rights of Full Members and those of countries that participate through Associate Membership or some other model. It is therefore possible for Canada to negotiate for full scientific and technological rights if an alternative to Full IGO Membership is sought.  Given the significant scientific leadership potential of the Canadian community as well as the potential for key technical contributions that deliver return on SKA1 construction funding investment (\S\ref{sec:Canada}), this approach aligns well with Canadian ambitions.


\vs

\begin{mdframed}[style=recstyle]
{\bf Recommendation 3. The membership model through which Canada participates in the intergovernmental organisation (IGO) that will build and operate SKA1 should provide full scientific and technological access as well as leadership rights for Canadian researchers and industry.}  An agreement for Canadian participation in the IGO should be finalized as early as possible in the next decade in order to maximize our impact on the construction phase as well as to maximize opportunities for technological tender and procurement.
\end{mdframed}

\newpage

\section{Responses to Panel Questions \label{sec:boxes}}

\vs\vs\vs

%





\begin{lrptextbox}[How does the proposed initiative result in fundamental or transformational advances in our understanding of the Universe?]

 Among the wide array of transformational advances in our understanding of the Universe that SKA1 will make (SKAO+ 15), those highlighted below align best with Canadian strengths (see Box~3 and Spekkens+ 19).

\no {\bf Pulsars:} Through wide-area pulsar search surveys and precision timing, SKA1 will discover extremely relativistic systems with which to test the predictions of general relativity as well as Cosmic Censorship and the No-hair theorem in black holes (Shao+ 15). SKA1 will also detect and characterize low-frequency gravitational radiation from supermassive black-hole binary systems (Janssen+ 15).  

\no {\bf Cosmic Magnetism:} SKA1 will perform an all-sky rotation measure (RM) survey of compact polarized background sources with a source density nearly three orders of magnitude higher than current datasets (Oppermann+12).  This RM grid will be a powerful probe of foreground magnetic fields on many scales and at many redshifts, from the distant Universe to the Milky Way itself (Johnston-Hollitt+ 15). 

\no {\bf Transients:} The combination of SKA1-Mid sensitivity and resolution could reveal FRB origins and emission mechanisms, detecting bursts up to $\sim$5 times fainter than current facilities and simultaneously providing precise ($\sim0.3$") localizations (Kaspi+ 19). Deep observations at $\sim5\,$GHz will connect the physics of accretion disks and relativistic jets (Corbel+ 15) and will characterize sources that are barely detectable today. 


\no {\bf Low-frequency Cosmology:} SKA1-Low will directly image the Epoch of Reionization (EoR) and Cosmic Dawn (CD) at $6 < z < 28$ on arcminute to degree scales (Koopmans+ 15). These earliest periods of the Universe's history accessible to SKA1-Low are uncharted, while tomography of CD/EoR elucidates the nature and evolution of the first luminous objects (Mellema+ 15). 


\no {\bf Galaxy Evolution:} SKA1-Mid will be the first facility to map the atomic hydrogen distributions in galaxies over cosmic time, probing the evolution of the galactic baryon cycle and angular momentum assembly (Blyth+ 15). Deep radio continuum observations will also probe faint radio halos of galaxy clusters both near and far (Cassano+ 15) to constrain how they form and evolve.

\no {\bf Multi-messenger Astronomy:} $1-10\,$GHz radio monitoring of the electromagnetic counterparts to GW events sheds important light on neutron star merger physics and the interaction between kilonova ejecta and environments (Abbott+ 17). As GW detectors improve in sensitivity, deeper radio follow-up will be required (Corsi+ 19). This will place SKA1 at the forefront of multi-wavelength, multi-messenger astrophysics.

\no {\bf Planetary System Formation:} SKA1 will play a crucial role in understanding protoplanetary disk physics in regimes that are inaccessible at millimeter wavelengths: the efficiency of dust and planetesimal growth from 0.1--5$\,$AU from host stars will be probed by SKA1 at $\sim10\,$GHz (Testi+ 15), and SKA1 will produce the first images of disk winds to constrain disk dissipation processes (Pascucci+ 18).  
\end{lrptextbox}


\begin{lrptextbox}[What are the main scientific risks and how will they be mitigated?]


The anticipated scientific returns from SKA1 are among the broadest of any current or planned facility (SKAO+ 15; see also \S\ref{sec:Canada} and Box~1). There are unknowns in every scientific field and open question that translate into scientific risks in survey or observation design. In aggregate, however, the science return on SKA1 will be transformational even if any one of its many scientific goals aren't achieved. With an unmatched combination of sensitivity, angular resolution, imaging quality and frequency coverage at metre and centimetre wavelengths for years to come (Fig.~\ref{fig:sens}), the potential for serendipity with SKA1 is also high (Harwit 81). The overall scientific risk of SKA1 is therefore low.
\end{lrptextbox}


\begin{lrptextbox}[Is there the expectation of and capacity for Canadian scientific, technical or strategic leadership?] 


Canada is poised to play SKA1 leadership roles in the following key scientific and technological areas:

\no {\bf Pulsars:} Canada is home to world-leading radio pulsar groups that attract researchers from around the globe. Canadians have a long history of leadership in pulsar search surveys and long-term timing campaigns (Fonseca+ 19, Stairs+ 19), including the ongoing NANOGrav collaboration. There is excellent potential for Canadian leadership in a KSP to test gravity using pulsars.

\no {\bf Cosmic Magnetism:} As long-standing world leaders in cosmic magnetism (West+ 19), Canadians are at the helm of almost all major ongoing or pending worldwide radio polarisation surveys, including those on SKA Pathfinders; there is also heavy Canadian involvement in radio surveys with a significant polarisation component. There is a strong expectation that Canada will be in a position to lead a KSP on cosmic magnetism.

\no {\bf Transients:}   Canadian expertise in transient studies is rapidly growing. Many Canadians have become world leaders in FRBs due to CHIME, and there is also significant expertise in accreting compact objects (Kaspi+ 19). A Canadian currently co-chairs the transients SKA Science Working Group (SWG), and Canadians are highly likely to be in a position to lead transient-focussed KSP or PI programs on SKA1.

\no {\bf Low-frequency Cosmology:} Canada is uniquely positioned to play a leading role in achieving the cosmology goals of SKA1-Low.  There is significant Canadian leadership within key SKA Precursors and other closely related experimental efforts.  Canadian expertise spans Cosmic Dawn through structure formation, measurements of fluctuations and globally averaged signals, and theory and experiment (Liu+ 19a, 19b).

\no {\bf Galaxies, Multi-messenger Astronomy, Planetary System Formation:} Canadian multi-wavelength expertise in galaxy evolution, multi-messenger astronomy and planetary system formation – in which radio observations play a critical role – is a key strength. Among other contributions, Canadians are involved in a variety of SKA Pathfinder and SWG initiatives in galaxy structure and evolution, have played important roles in the discovery and multi-wavelength follow up of GW170817 and  formed a coalition for multi-messenger astrophysics (Ruan+ 19), and have a strong exoplanet community (Benneke+ 19) and expertise in protoplanetary disk studies (Matthews+ 19) that could be  extended to the longer wavelengths probed by SKA1.

\no {\bf Technology:} Canada has played a major role in technology development for SKA throughout the design phase, and can anticipate a similar leadership role during construction and operations (Rupen+ 19; Spekkens+ 19). There is long-standing Canadian expertise in areas such as digital signal processing (e.g., correlators and beamformers),  combined  with  a  long  history  of  successful,  major  contributions  to  large  international  observatories like ALMA and Gemini.   Canada has provided the baseline design for the SKA1-Mid correlator/beamformer which forms the heart of the telescope, as well as major elements of the dish design; Canada also leads SKA1 work on software interfaces and analysis of the signal chain, both essential aspects of the project.

\no {\bf Compute:} Scientific computing platform and archive development is a Canadian strength (Kavelaars+ 19), and the Canadian Advanced Network for Astronomical Research (CANFAR) as well as the Canadian Astronomy Data Centre (CADC) could play leading roles in the development of SRC architectures and implementations. Some of the capabilities that will be needed for a Canadian SRC are being developed now through the Canadian Initiative for Radio Astronomy Data Analysis (CIRADA; West+ 19) that will produce, archive and visualise science-ready data products from ASKAP, CHIME and the (J)VLA Sky Survey (VLASS). 


\end{lrptextbox}


\begin{lrptextbox}[Is there support from, involvement from, and coordination within the relevant Canadian community and more broadly?] 


Canada has been scientifically and technologically engaged in the SKA initiative for over two decades (e.g.\ Ekers 12, Spekkens+ 19). Examples of ongoing scientific involvement include Canadian membership in every SKA Science Working Group and leadership in SKA Pathfinder surveys and instruments (see Box~3). Current Canadian technological involvement exploits industry partnerships to play leadership roles in SKA1 R\&D, the long-standing collaboration between MDA and NRC in Central Signal Processing being a prime example (see Box~3). Project coordination within the Canadian community is the responsibility of the ACURA Advisory Committee on the SKA (AACS) and the Canadian SKA Science Director, who reguarly communicate with the broader   community.
Coordination between Canada and the global SKA community is enabled through participation in SKAO science (e.g.\ the Science and Engineering Advisory Committee), technology (e.g.\ the SKA Regional Centre Steering Committee) and governance (e.g.\ the Board of Directors) committees. Canadian SKA activities are well-supported and well-coordinated both nationally and internationally.
\end{lrptextbox}


\begin{lrptextbox}[Will this program position Canadian astronomy for future opportunities and returns in 2020-2030 or beyond 2030?] 

\no {\bf Science:} A construction start date for SKA1 of Q2~2021 implies that most of its scientific opportunities and returns will materialize beyond 2030 (Fig~\ref{fig:timeline}); this is particularly the case for the fundamental advances highlighted in Box~1, many of which likely require KSP-like allocations to achieve (Braun+ 15). Beyond the scientific avenues provided by SKA1 alone, the competitive advantage gained through access to the dominant metre and centimeter facility on the planet will position Canadians for future opportunities beyond these wavelengths. Since Canadians already excel in multi-wavelength/multi-messenger astronomy, the community is likely to leverage its SKA1 access effectively. The obvious synergies between SKA1 and other major initiatives are being recognized at the facility level, which may result in interfacility agreements that benefit Canadians. For example, the SKA and ngVLA are already exploring the possibility of an alliance to provide partners in one facility with access to the other.

\no {\bf Technology:} SKA-related R\&D has had significant knock-on benefits for Canadian astronomy (Rupen+ 19).  For example, the composite dish technology developed for SKA1 was not ultimately selected for the Baseline Design, but has made Canada a world leader in cutting-edge dish design at both low and high frequencies; this approach has been adopted in both the CHORD (Vanderlinde+ 19) and ngVLA (Di Francesco+ 19) designs.  Similarly, the TALON-DX processing board developed for the SKA1-Mid correlator has kept Canada at the forefront of international correlator design, as evidenced by strong interest in that technology from other large interferometers like ALMA and the ngVLA.  Further, the archiving and data access techniques developed for SKA1 signal processing and digital research development in the SRC context have obvious applications to ``big data'' within and beyond astronomy (Kavelaars+ 19).  Engaging with SKA1 will maintain and enhance Canada's reputation as an excellent technological partner in future national and international initiatives.

\end{lrptextbox}


\begin{lrptextbox}[In what ways is the cost-benefit ratio, including existing investments and future operating costs, favourable?] 


\no {\bf Benefits:} SKA1 will be the dominant general-purpose observatory at centimeter and meter wavelengths for years to come, 
and the scientific benefits are correspondingly high across a broad range of fields (Boxes~1~3,~5). Technological participation in SKA1 will also yield a range of benefits and future opportunities (Box~5).

\no {\bf Cost-benefit ratio:} Canadian signature SKA1 technologies are the key mechanism for ensuring a favourable cost-benefit ratio for SKA1 participation. Canada is a leader in SKA R\&D through effective partnerships between academia and industry (Box~5; Rupen+ 19; Spekkens+ 19).  Our key SKA1 technological capabilities include the design and fabrication of correlators and beamformers, digitisers, low-noise amplifiers, signal processing, and monitor \& control.  These technologies provide a suite of possible in-kind contributions to offset construction costs for good return on construction investment. Initial estimates suggest that the fair work return across all SKA1 construction partners will be high (C\'esarsky 19b) . 

\no {\bf Future operating costs:} The best available cost estimate for a 6\% participation in SKA1 during full operations ($>2030$) is ($\sim$\$10M operations + $\sim$\$5M SRC  =) $\sim$\$15M/year (2017~CAD), but this number is very uncertain. These operating costs 
are consistent with those estimated for the ngVLA (within the considerable uncertainties; Di~Francesco+ 19) and ALMA, the other astronomical interferometers of similar scale.

\end{lrptextbox}


\begin{lrptextbox}[What are the main programmatic risks
and how will they be mitigated?] 

\vsneg\vsneg\vsneg

Four main programmatic risks for SKA1 are described below; the first three are project-wide, and the last one pertains specifically to Canada.  The mitigation strategies described focus on Canadian options for retiring risk.

\no {\bf 1.\ Milestones before construction:} The construction start date of Q2~2021 (Fig.~\ref{fig:timeline}) requires that key project milestones occur over the next year.  Those for SKAO include System CDR, Cost Book development as well as construction and operations proposal preparation. Those for the IGO include Treaty Convention ratification, procurement policy development and construction funding approval. A delay in any milestone could lead to a slip in construction start. While some milestones (e.g. Cost Book, procurement policy) are within the purview of the SKAO/IGO, others (e.g.\ Treaty ratification, funding commitments) are largely beyond the project's control. A slip in construction start is costly, and could affect performance if countries dip into construction funds to close the gap (see risk 2). At present, Canada can only help mitigate SKAO milestone risks (see risk 4). Canadian engagement and leadership are particularly important as SKAO produces key deliverables and winds down as an organisation, and we are well-positioned to contribute (Box~4 and Spekkens+ 19).  

\no {\bf 2.\ Building the Design Baseline:} The SKA1 Design Baseline is mature and its technical risk is relatively low, but an externally reviewed Cost Book is forthcoming only in Q2~2020. The estimated cost of Design Baseline construction and operations (\S\ref{sec:project}) therefore remains uncertain as the design phase comes to an end. If the final Design Baseline cost greatly exceeds available funding at the start of construction, then the facility that is initially built (the Deployment Baseline) could be significantly less capable than required to deliver the advances in Box~1. The scalable nature of radio interferometers and the SKA1 design approach imply that the Design Baseline can be recovered as more funding becomes available, but that delay incurs costs (risk 1) and risks the loss of low-hanging scientific ``fruit" to other facilities. Risk mitigation for Canada includes scientific and technical engagement in Deployment Baseline development in Q1-Q3~2020, for which we are well-positioned (Box~4 and Spekkens+ 19). If the Q3~2020 Deployment Baseline differs significantly from the Design Baseline, Canada may need to re-evaluate its SKA1 ambitions.  


\no {\bf 3.\ The SRC compute model:} In the SKA1 SRC model, scientific processing and storage as well as observing and archive user support are funded separately from construction and operations by a network of external facilities (\S\ref{sec:project}). This poses the risk that the scientific exploitation of SKA1 is hindered by a lack of global SRC capacity rather than by its technical capabilities, a challenge already being faced by some SKA Pathfinders. The SKAO is investigating the possibility of including minimal science compute within the operations budget in order to mitigate some of this risk, and Canada can contribute through continued leadership in defining SRC requirements and costs (\S\ref{sec:Canada} and Box~3). Hosting a Canadian SRC that includes processing, storage and user support tailored to our community needs would also retire some of this risk.  


\no {\bf 4. Canadian IGO participation:} Canada's Observer status on the Council Preparatory Task Force (CPTF) of the IGO is an important step towards participating in SKA1 construction and operations. However, there is currently no mechanism for Canada to inform policy or process during the rapidly approaching construction and operations phases (\S\ref{sec:project}), and the IGO participation model for Canada remains undefined. The  programmatic risk is twofold. First, Canada risks losing agency in SKA1 if the SKAO winds down before our IGO participation is finalized, forfeiting the opportunity to provide scientific and technical leadership during that time. Second, Canada risks missing out on construction tender and procurement if no participation mechanism exists at that time, which would significantly increase the cost-benefit ratio of SKA1 participation (Box~6). These risks can be mitigated if an agreement for Canadian participation in the IGO is finalized early in the next decade.

\end{lrptextbox}

\vsneg\vsneg

\begin{lrptextbox}[Does the proposed initiative offer specific tangible benefits to Canadians, including but not limited to interdisciplinary research, industry opportunities, HQP training,
EDI,
outreach or education?] 

\vsneg\vsneg

Canadian participation in SKA1 will provide tangible benefits to industry early in the next decade (Boxes~3,~5,~6), while scientific leadership will thrust astronomers into the global limelight later on (Boxes~1,~3,~5). Both will attract top HQP talent to Canada across a range of disciplines, will grow and sustain our world-leading Canadian astronomical community, and will provide inspiration for future generations of Canadian scientists.

\end{lrptextbox}

\end{document}